# BCMI-Driven Motion Control Detection: EEG-Based Machine Learning and Interaction Entropy for High-Order Brain Networks

Jiajia Li[1], Fan Li[1], and Jian Song[2*]

[1] College of Information and Control Engineering, Xi'an University of Architecture and Technology, Shaanxi, Xi'an 650500, China
[2] Department of Neurosurgery, General Hospital of Central Theater Command, Wuhan 430070, China
Correspondence e-mail: songjian0505@smu.edu.cn (JS)

**Abstract:** This study investigates the cognitive motor control detection and the underlying neuroregulatory mechanisms during music-assisted simulated driving. Using a dynamic higher-order network model constructed with EEG-based cross-information entropy, we quantify the dynamic coordination within brain networks activated during both music listening and driving. This approach, which contrasts with previous static network analyses, provides novel insights into how musical stimuli modulate the complex interplay of brain regions during demanding tasks. Results demonstrated enhanced third-order connectivity and elevated higher-order information entropy in music-stimulated driving compared to baseline driving, as evidenced by increasing Φ values of higher-order network indices. Supervised machine learning, including support vector machines, revealed a strong correlation between model accuracy and ROC-AUC values and the hierarchy of brain network features. This underscores the importance of higher-order features in decoding brain motor-control states during music-simulated driving. These findings deepen our understanding of the interplay between music cognition and motor control, offering valuable insights for the development of novel brain-computer-music interfaces (BCMI) and adaptive human-machine systems to enhance performance in demanding tasks like driving.

## I. Introduction

Brain-Computer Music Interface (BCMI) have garnered substantial attention in both neuroscience and applied domains. Early endeavors, such as Barrett's 1965 work on PDR-based percussion control [1], laid the foundation, while modern BCMIs have expanded beyond brain-generated music to encompass the regulation of cognitive functions—including emotion and motor control [2], [3], [4]. Against this backdrop, the present study focuses specifically on investigating the regulatory effects of musical melodies on motor execution.

Research in motor cognition has identified key determinants of decision precision, namely stability maintenance [5], [6], [7], motor preparation [8], [9], [10], and dynamic control [11], [12], [13]. Environmental factors further modulate such processes: open environments, for instance, facilitate the execution of motor strategies [14], while visual stimuli optimize motor decision-making in rodents [15], [16]. Among such environmental influences, music-based interventions have been shown to enhance motor intelligence through attentional mechanisms [17], [18] and neural synchronization [19], with accumulating evidence suggesting the involvement of higher-order network modulation as well [20].

Notably, music exerts these effects by engaging multisensory processing—encompassing visual, motor, and auditory modalities—thereby increasing network complexity during multitasking. This distinguishes it from other sensory inputs: while tactile-motor integration optimizes mechanical behavior [21] and visual cues enhance decision precision [22], [23], music outperforms non-musical auditory stimuli in optimizing visual-motor efficiency [24], [25], a superiority potentially rooted in higher-order network integration.

Neuroscientifically, these effects of music on brain function are underpinned by specific oscillatory mechanisms: theta oscillations (4–8 Hz) mediate syntax processing and emotional valuation [19], [26]; beta-band connectivity (14–30 Hz) predicts rhythmic processing and motor learning [27], [28]; gamma power (31–45 Hz) correlates with information entropy [29]; and alpha modulation (8–13 Hz) directs attention with high spatial specificity [30], [31], [32].

Beyond oscillatory mechanisms, music further reorganizes multiscale network topology, as evidenced by shifts in current density during tinnitus treatment [33], enhanced inter-brain synchronization [34], beta-band connectivity changes influencing memory encoding [35], and the shortening of prefrontal-striatal pathways during emotional processing [36].

However, current methodologies struggle to capture these complex neural dynamics, being constrained by pairwise linear metrics (e.g., phase locking value) that fail to account for multi-node nonlinear synergies [26], [33], alongside static network analyses that overlook dynamic reconfiguration processes [34], [37]. To address these limitations, this study integrates phase synchronization with hypergraph theory [39] and machine learning [38] to decode the mechanisms underlying music cognition.

Through structural and functional module analysis, the proposed approach aims to: (1) quantify global phase synchronization using node degree and segregation-integration metrics; (2) characterize high-order network reorganization by leveraging Φ-entropy within hyperedges; and (3) validate the findings through music-enhanced motor-visual tasks, utilizing navigation bias angle metrics.

## II. MATERIALS

### A. Participants

This study enrolled 20 right-handed undergraduate or graduate students (24 ± 1 years) with normal or corrected-to-normal vision. The research protocol was approved by the Ethics Committee of the General Hospital of Chinese PLA Central Theater Command (reference number: [2020]041-1). Participants were free from color blindness, allergies, asthma, hypertension, and neurological disorders. Written informed consent was obtained from all participants. They were instructed to abstain from alcohol, caffeine, and other central nervous system stimulants for 48 hours prior to the experiment and to maintain clean scalps for optimal EEG signal acquisition. All participants achieved proficiency in operating the simulated driving platform following practice.

### B. Experimental Design

This study was carried out in the Campus Information and Control Building. A human-computer interaction driving simulation system was established using the City3D game engine [see Fig. 1(a)], comprising a simulation platform with a clutch, brakes, accelerator, gearshift, and a steering wheel equipped with a mobile device. The PC ran the City3D game, featuring an urban road environment to simulate real-world driving conditions [40], [41], [42], [43], [44], [45].

The experiment consisted of three phases: Resting State (RS), Baseline Driving (BD), and Music Stimulated Driving (MSD), with a total duration of 12 minutes [see Fig. 1(b)]. The RS phase lasted 3 minutes, during which participants remained awake without performing tasks. The subsequent 9 minutes involved simulated driving, divided into BD (6 minutes) and MSD (3 minutes). Following RS, participants initiated driving on a straight road under researcher instruction, entering the BD phase. After 6 minutes, music stimulation was transmitted via a music delivery device, transitioning the experiment to the MSD phase. Upon completion, participants stopped the vehicle under researcher instruction.

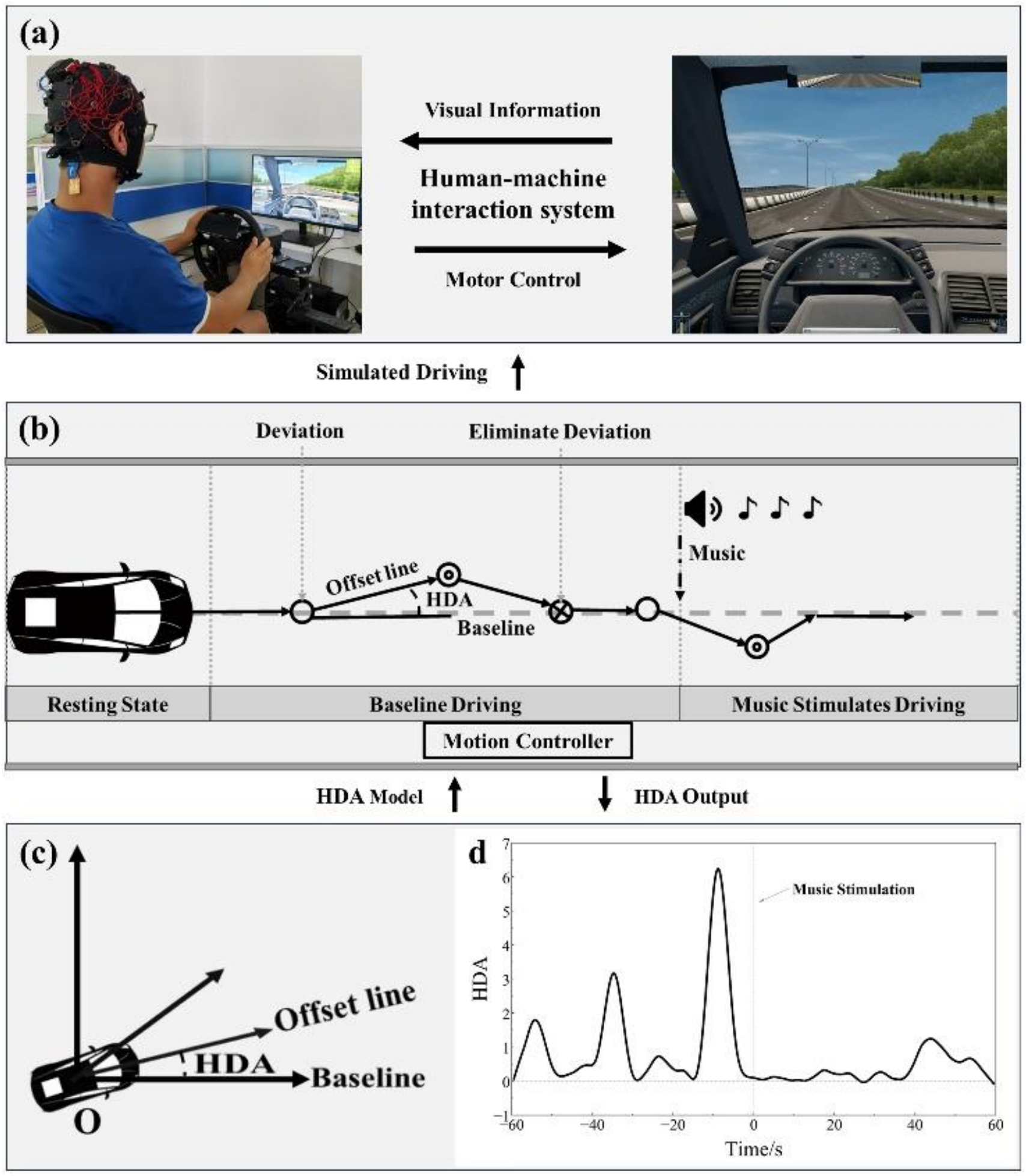

**Fig. 1.** Conceptual framework of the experimental paradigm. (a) A Human-Computer Interaction (HCI) environment where participants receive visual information through a simulated driving system while providing motor control feedback, such as steering wheel manipulations. (b) A simulated driving task paradigm consisting of three sequential stages: resting state (RS), baseline driving (BD), and music-stimulated driving (MSD). (c) An analytical model depicting the measurement of course deviation angles during the simulated driving tasks. (d) A comparative time series analysis of driving HAD (representing real-time motor-control level)

between the BD and MSD stages.

During simulated driving, the reference line was defined by the steering wheel's center position and its forward extension, while the deviation line was determined by the steering wheel's center and the top position during movement [see Fig. 1(c)]. The angle between these lines was recorded as the Heading Deviation Angle (HDA) [40]. A mobile device mounted on the steering wheel collected data, requiring participants to adjust the wheel to maintain road centerline alignment. Prior to the experiment, participants practiced for 5-10 minutes to ensure proficiency and comfort with the simulation platform. Brain activity and music transmission devices were fitted post-practice. The experimental environment was maintained at a comfortable temperature with soft lighting and minimal noise. This study adhered to the Helsinki Declaration and ethical standards, with all participants providing informed consent.

This study focused on behavioral changes during BD and MSD phases, as the RS phase did not involve driving tasks. As shown in Figure. 1(d) music stimulation significantly reduced the offset of steering angle deviation time curves, indicating enhanced driving stability. This improvement in driving stability can be attributed to music's impact on cognitive resource allocation, attentional modulation, and emotional feedback mechanisms.

### *C. Data Collection and Preprocessing*

The EEG data were collected using a Biosemi 32-channel amplifier (Neuroscan) with Ag/AgCl electrodes mounted on an elastic cap according to the international 10-20 system, using CMS-DRL as the reference point. The signals were recorded from 32 electrodes with a raw signal sampling rate of 1024 Hz, which was subsequently down-sampled to 128 Hz. This ensures the offset relative to the CMS-DRL reference remained below 20 μV. The EEG data preprocessing involved the following steps:

1) *Channel localization:* Identified all channel information and assessed the validity of each electrode signal.
2) *Re-referencing:* Re-referenced all channels to the computed average reference.
3) *Filtering:* Applied a basic FIR filter with a high-pass of 0.5 Hz and a low-pass of 45 Hz.
4) *Artifact removal:* Removed artifacts (e.g., eyeblinks, eye movements, and head movements) using Independent Component Analysis (ICA).
5) *Segmentation:* Segmented the preprocessed EEG signals into three stages (RS, BD, and MSD) based on the experimental tasks for further processing and analysis.

In this study, a mobile device featuring an in-house developed angle measurement app was mounted at the steering wheel's center to continuously monitor HDA throughout the driving task. The HDA data underwent a three-step preprocessing protocol:

1) *Filtering:* A Butterworth low-pass filter was applied to minimize high-frequency muscle artifact noise during driving.
2) *Data Fitting:* Cubic spline interpolation was performed on the filtered data to enhance local feature representation and temporally align the steering angle deviation with EEG recordings.
3) *Baseline Correction:* A polynomial-based baseline estimation was conducted on the fitted data, followed by baseline subtraction from the original time-response curves to yield baseline-corrected data for subsequent analyses.

## III. High - Order Brain Network Features Based on PLV

**TABLE I** ROI regions vs. EEG channels

| ROI Regions | Channels |
|---|---|
| DMN | P3, P4, Pz, PO9, PO10 |
| SMN | C3, C4, FC5, FC6, CP5, CP6, P7, P8 |
| VN | O1, Oz, O2 |
| AN | T7, T8, FT9, FT10 |

A high-order brain network algorithm was developed in this paper, which integrates the analyzing features of PLV and mutual information. The proposed algorithm examines how music influences high-order brain network properties during driving. Fig. 2(a) outlines the algorithm's workflow: raw EEG data undergo preprocessing, followed by ROI selection (Table I) focusing on key networks - default mode network (DMN), sensorimotor network (SMN), visual network (VN), and auditory network (AN). This ROI-based approach enhances analytical efficiency while avoiding the complexity of whole-brain analysis.

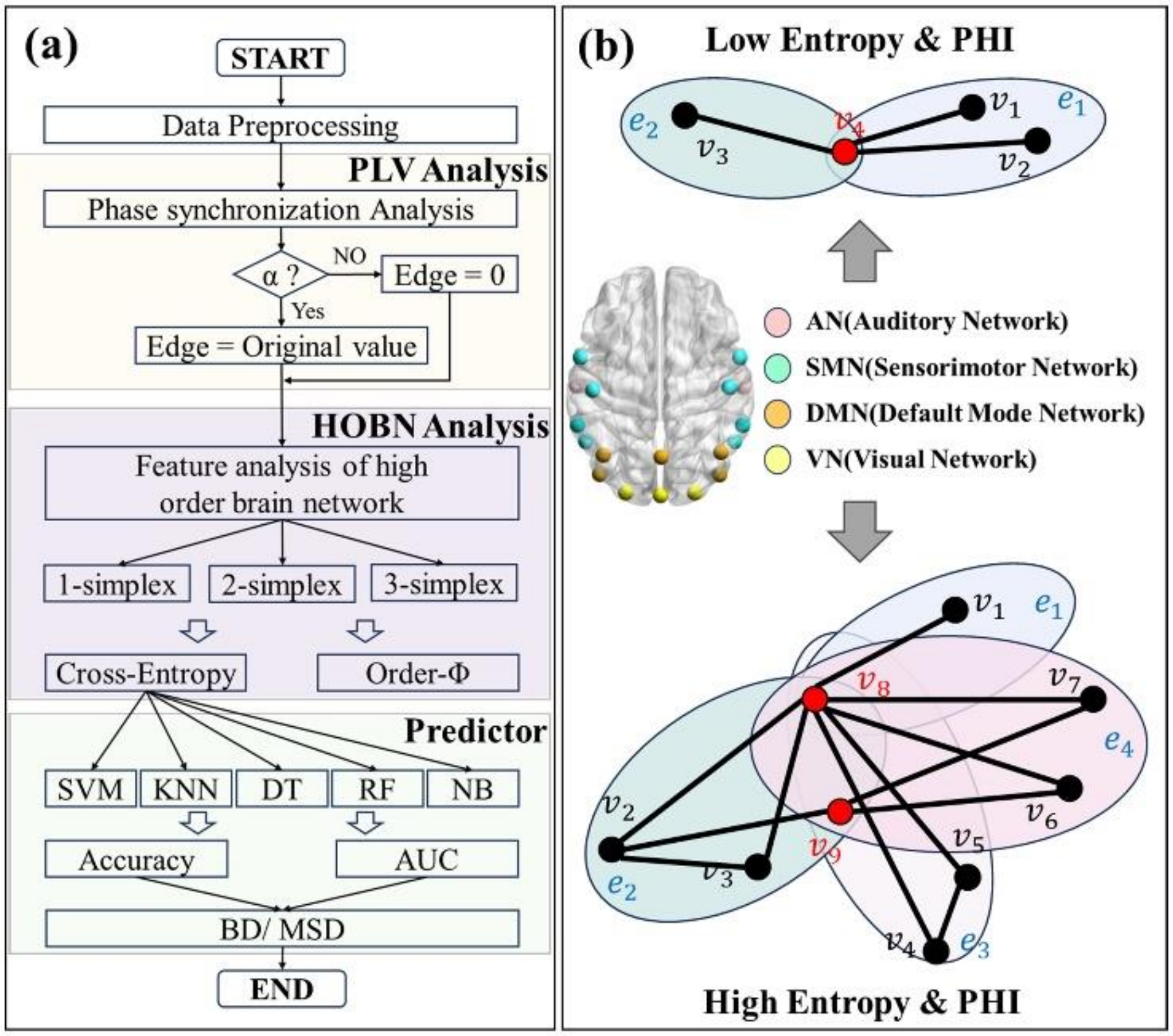

**Fig. 2.** Illustration of high-order network feature analysis. (a) Flowchart of the developed brain state decoding algorithm utilizing high-order network feature analysis. (b) Distribution of network structures with varying interaction entropy and integrated information Φ.

Phase synchronization analysis was conducted on ROI data to generate synchronization matrices for RS, BD, and MSD stages. We focused on BD and MSD stages for in-depth analysis, as RS represented a task-free resting state. The threshold parameter α was defined as 0.3 times the maximum value of the difference matrix (*diff*) between BD and MSD. Elements x in the BD and MSD matrices were retained if $x \geqslant \alpha$; otherwise, they were set to zero. This thresholding process identified edges with significant differential changes, referred to as key edges.

Then, using higher-order network theory, interaction entropy and Φ values were computed for key edge matrices across 1-simplex, 2-simplex, and 3-simplex networks in both BD and MSD conditions. To assess the predictive capability of higher-order network metrics for brain working modes, we implemented five supervised machine learning models: Support Vector Machine (SVM), K-Nearest Neighbor (KNN), Decision Tree (DT), Random Forest (RF), and Naive Bayes (NB). Prediction accuracy and AUC values were computed to evaluate the classification performance of higher-order metrics in distinguishing brain working states

To evaluate the complexity, stability, and efficiency of information transmission within the network, as well as its capacity for information integration, we computed interaction entropy and Φ values derived from higher-order network analysis in prior studies. As illustrated in Fig. 2(b), networks with elevated interaction entropy and Φ *v*alues exhibited more refined information transmission and broader hyperedge participation. Conversely, networks with lower values demonstrated reduced information transmission and narrower hyperedge coverage.

The interaction entropy of higher-order networks represents a generalization of traditional Shannon entropy, specifically designed to quantify system complexity through the probabilistic properties of hyperedge distributions in higher-order network structures.

Let $H$=($V$,$E$) be a hypergraph, where E forms a simplicial complex. Let e denote a probability measure defined on the hyperedge set E. The interaction entropy is then defined as:

$$S_H = -\sum_{e \in E} P(e) log P(e) \tag{1}$$

Where $P(e)$ can be calculated based on hyperedge weights and their occurrence frequency.

The Φ value represents a higher-order node centrality measure that quantifies node influence in higher-order interactions. While traditional centrality metrics focus on individual node properties, the Φ value assesses node importance based on hyperedge coverage and overlap. Specifically, hyperedge coverage denotes the extent to which each hyperedge a node participates in covers other nodes, while overlap weight reflects the degree of overlap a node has across different hyperedges. The Φ value is formally defined as:

$$\Phi(v) = \sum_{e \ni v} \frac{1}{|e|-1} \sum_{u \in e\{v\}} \frac{1}{\deg(u)} \tag{2}$$

Where $|e|$ represents the hyperedge size, and deg(u) represents the node degree.

In neuroscience and cognitive science, interaction entropy and Φ values are pivotal metrics. They offer complementary perspectives on brain complexity, dynamically tracking changes in brain states. While interaction entropy quantifies information uncertainty and distribution, Phi values capture a system's capacity for information integration. These measures not only provide insights into the neural substrates of cognition but also aid in both the research and diagnosis of neurological disorders, furnishing critical quantitative evidence and guiding research into brain mechanisms, disease diagnosis, and treatment.

## IV. Result

### *A. Brain Topological Patterns Based On PLV Analysis*

In this study, we modeled EEG channels as network nodes and constructed brain functional networks using PLVs to quantify inter-channel phase synchronization. To investigate how simulated driving with and without music affects brain function, we systematically analyzed phase synchronization patterns across task stages [see Fig. 3(a)].

Transitioning from RS to BD engages the frontal cortex as a high-order cognitive control hub, enhancing internal functional connectivity to coordinate driving-related activations. The central sensory-motor cortex undergoes neural reorganization, showing strengthened phase synchronization with the frontal cortex, indicative of optimized cognitive-motor pathways. The limbic system, serving as an auxiliary network, intensifies information exchange with core brain regions to support initial task integration demands.

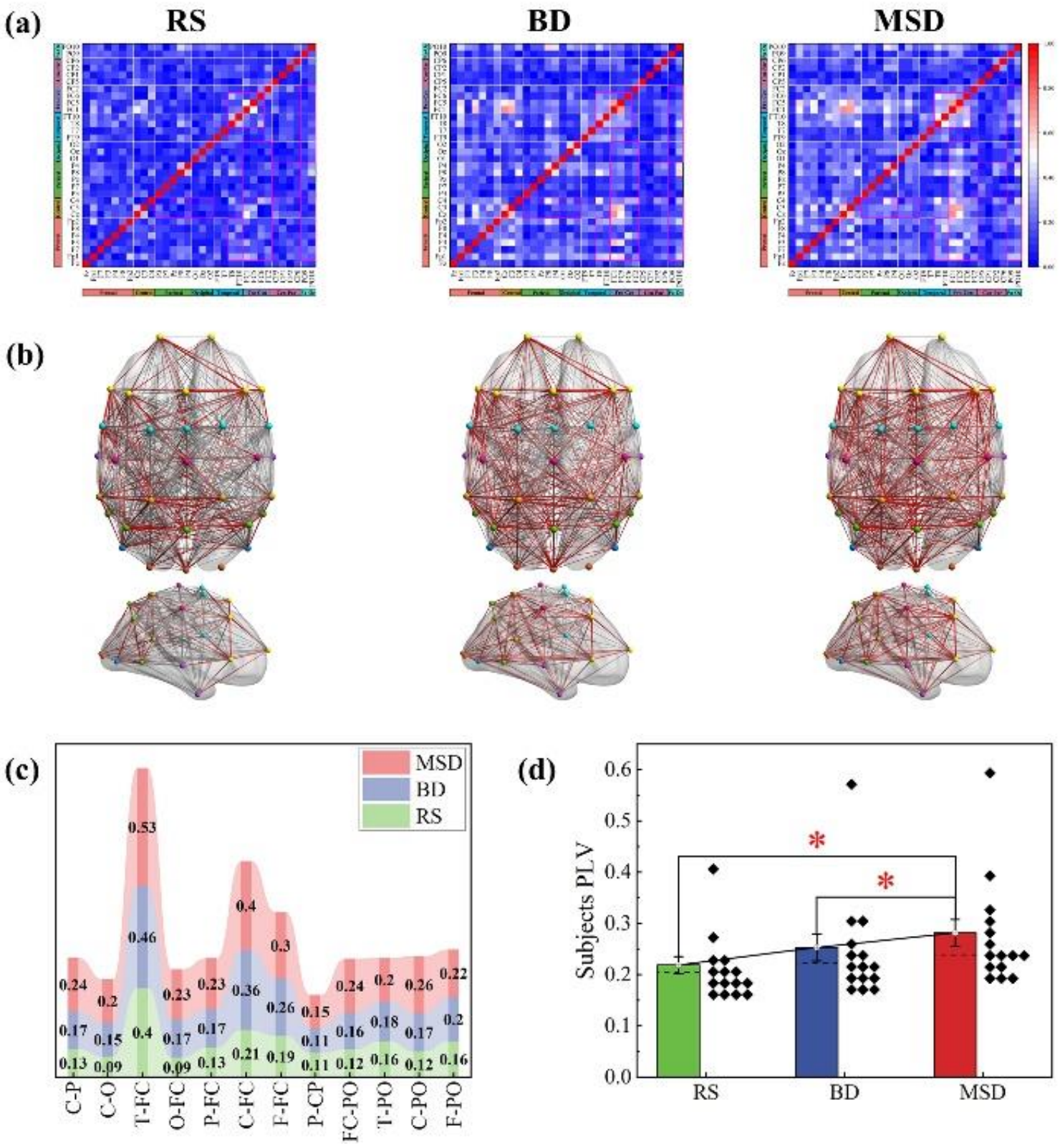

**Fig. 3.** The evolution of brain network topology across RS, BD, and MSD stages. (a) Phase synchronization matrices corresponding to different task stages. (b) Brain network topological patterns at different task stages. (c) Comparison of mean PLV values among brain regions across different task stages. (d) Statistical significance of mean PLV values across subjects and task stages. *: $p$<0.05; **: $p$<0.01.

During MSD stage, the brain exhibits multitasking characteristics. The frontal cortex maintains driving control while establishing functional connectivity with the temporal auditory processing area, enabling parallel processing of musical information. The frontal medial region enhances neural synchronization with frontal, central, and parietal regions, forming a dynamic regulation mechanism that integrates motor precision with multimodal decision processing. Notably, cortico-cortical interactions between parietal and occipital regions significantly enhance central synchronization, suggesting deep integration of spatial navigation with visual-motor information under dual-task conditions.

Topological analysis of the brain network [see Fig. 3(b)] confirms these findings. Compared to RS, BD shows increased whole-brain connection density, particularly in frontal-central sensory-motor network connectivity, reflecting cognitive-motor integration demands. In MSD, whole-brain connection density further increases, with the frontal medial region forming a core control hub through enhanced premotor cortex connectivity, enabling refined behavioral control. Increased parieto-occipital connectivity suggests spatial attention reallocation under multitasking. As shown in Fig. 3(c), fronto-temporal pathway synchronization is enhanced, indicating dynamic cognitive control over auditory processing. Strengthened parieto-occipital-temporal synchronization enables rhythmic information to map onto visual-motor representations in the parietal cortex. Enhanced fronto-parieto-occipital connectivity reflects dynamic attention allocation, prioritizing critical visual processing in the parieto-occipital cortex while suppressing irrelevant information. Music enhances this selective attention through its regulatory effects.

Fig. 3(d) displays the alterations in the mean phase synchronization matrix across 15 participants during the task phase. Compared to the BD phase, the whole-brain phase synchronization mean exhibited a significant increase during the MSD phase (paired sample t-test: $t$=5.072, $df$=14, $p$<0.05), accompanied by reduced interparticipant variability. This indicates that enhanced interregional coordination demonstrates significant group-level consistency, suggesting that musical intervention can reliably augment regional coordination while improving core circuit interaction efficiency. These findings align with mechanisms underlying multimodal integration and resource optimization.

The brain network coordination mechanism in complex driving environments is mediated by the "dorsal medial prefrontal-temporal-parieto-occipital" circuit, which facilitates multi-level functional integration through three primary mechanisms:

*Dorsal medial prefrontal coordination enhancement:* Centered on real-time coupling between prefrontal decision signals and central sensory-motor control, this mechanism establishes a "perception-action" closed-loop system. Bidirectional connections enable precise motor execution coordination.

*Multimodal interaction mechanism:* Temporal lobe auditory processing and occipital lobe visual inputs converge in the parietal lobe, generating dynamically updated spatial-auditory-motor representations. The parieto-occipital region, as a cross-modal integration hub, incorporates musical rhythms and spatial road information into a unified driving context model.

*Resource competition regulation:* The prefrontal cortex suppresses non-task-related neural connections, enhancing coordination efficiency within the core "dorsal medial prefrontal-temporal-parieto-occipital" circuit.

This circuit operates under a hierarchical integration principle. In the cognitive control layer, led by the prefrontal cortex, the dorsal medial prefrontal cortex functions as the regulatory hub, coordinating resource allocation between sensory-motor networks and the limbic system via bidirectional connections, forming a "decision-execution" information flow loop. Multimodal auditory and visual information is integrated to generate multidimensional environmental representations, supporting real-time driving adjustments. The limbic system evolves into a multimodal integration platform, with dynamic coupling strength increasing in proportion to task complexity.

This coordination framework elucidates brain network reorganization in music-driving scenarios. Through global resource allocation in the dorsal medial prefrontal cortex, multimodal fusion in temporal-parieto-occipital regions, and motor execution in the central region, a closed-loop control system ("auditory rhythm analysis →spatial context modeling →motor control output") is established. This system represents a dynamic resonance pattern evolved by the neural system to address multi-task competition, providing a theoretical model for understanding cognitive resource allocation in complex driving environments.

**TABLE II** SIGNIFICANT ONE-SAMPLE T-TEST RESULTS FOR THE DIFFERENTIAL EDGES BETWEEN BD AND MSD

| | Test Value = 0 | | | | | |
|---|---|---|---|---|---|---|
| | | | | | 95% Confidence Interval for the Difference | |
| | t | df | p-value (2-tailed) | Mean Difference | Lower Bound | Upper Bound |
| * | 15.123 | 33 | <.001 | .12301 | .1065 | .1396 |

## B. Significant Connectivity Selection for the Music-assisted Motor Control

The human brain serves as the central regulatory hub, with its functional mechanisms representing a focal area of scientific inquiry. Various physiological processes, including perception, motion, cognition, and other functions, are mediated by the orchestrated operations of multiple neural networks. Neuroscientific research has identified several pivotal networks, such as the Default Mode Network (DMN), Sensory-Motor Network (SMN), Visual Network (VN), and Auditory Network (AN). These specific regions of interest (ROI) are essential for executing diverse brain functions, with their interregional interactions forming the basis for complex brain operations. Recent investigations have focused on identifying key connections between these ROI regions, aiming to delineate their connectivity patterns and uncover critical edges that play pivotal roles in brain information transmission and functional coordination. This research provides fundamental theoretical insights into elucidating the brain's operational mechanisms.

This study analyzed 20 nodal data, dividing them into four ROI regions: default mode network (DMN), salience network (SMN), ventral attention network (VN), and auditory network (AN). Phase-locking matrices were computed between these regions during RS, BD, and MSD stages. We focused on comparing the BD and MSD stages, as RS represents a task-free resting state. Using a threshold of $\alpha = 0.3 * max(\text{MSD - BD})$, we identified 34 edges with significant differential changes between the BD and MSD stages. A single-sample t-test (see Table II) on these 34 edges revealed a significant phase-locking difference (M = 0.1230, SD = 0.04743), with $t(33) = 15.123$, $p < 0.001$, $d = 2.59$, and a 95% confidence interval [0.1065, 0.1396]. This indicates enhanced phase synchronization in the MSD stage compared to BD under musical stimulation. These 34 edges served as key connections, facilitating information exchange between the four functional networks.

The differential matrix between BD and MSD [see Fig. 4(a)] highlights the distribution of these key edges, which are primarily located between SMN and DMN, VN and DMN/SMN, and AN and DMN/SMN/VN. This clear visualization underscores the role of these key edges in mediating interregional connections within the functional modules.

This study quantified the connectivity strength of key edges among the DMN, SN, VN, and AN. The results demonstrated significant enhancement in the average connectivity strength of key edges between these networks following music stimulation, particularly in the BD and MSD stages. This suggests that music facilitates real-time inter-regional information exchange via strengthened key edges, enabling efficient cognitive processing and precise motor control during complex tasks.

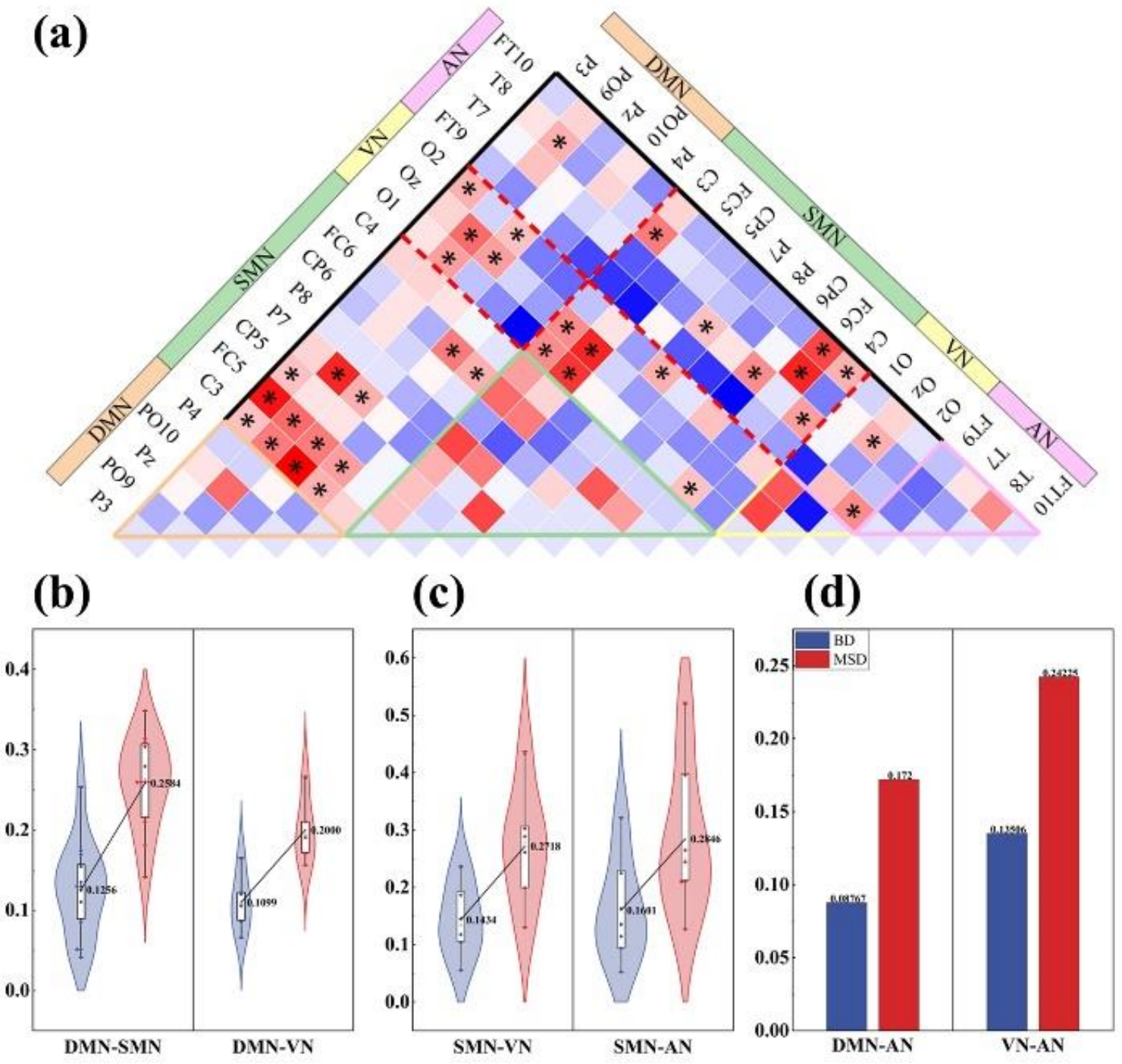

**Fig. 4.** ROI connectivity analysis between the BD and MSD tasks. (a) Distribution of network edges based on node-based connectivity matrices. (b) Box plot illustrating key ROI edge connection strength between DMN and SMN, VN. (c) Box plot showing key edge connection strength between SMN and VN, AN. (d) Comparative analysis of key edge connection strength between AN and both DMN and VN. * denotes key edges.

As illustrated in Fig. 4(c), music stimulation significantly strengthened multiple connectivity edges between AN and SMN, indicating enhanced neural coupling between auditory perception and motor responses. This improvement promoted action automation and coordination through the auditory-motor mechanism, ultimately enhancing driving control. Additionally, the strengthened connectivity between VN and SMN reflected improved efficiency in visual information processing and motor execution. Music stimulation optimized attention allocation by reducing the cognitive load associated with visual processing, thereby accelerating the conversion of visual information into driving actions through the VN-SMN "visual-perceptual motor" loop.

Fig. 4(b) shows that key edges between DMN and SMN were also strengthened, reflecting improved coordination between "planning-execution" processes during driving. While DMN is involved in high-level strategic planning, SMN manages specific motor operations. Enhanced DMN-SMN connectivity may therefore improve driving adaptability. Furthermore, strengthened VN-DMN connectivity suggests a balance between "internal-external" attention, as music stimulation maintained visual attention while preserving DMN's self-monitoring function, thereby optimizing cognitive resource allocation.

As shown in Fig. 4(d), the strengthening of key edges between AN and DMN was relatively modest but still evident under music stimulation. This may be attributed to the indirect influence of music on DMN, primarily through its interaction with SMN, as DMN is more involved in self-monitoring and self-adjustment rather than direct auditory perception. The relatively independent processing pathways of AN and VN in the brain, combined with reduced cognitive load in visual processing under music stimulation, resulted in a stable interaction pattern between these networks.

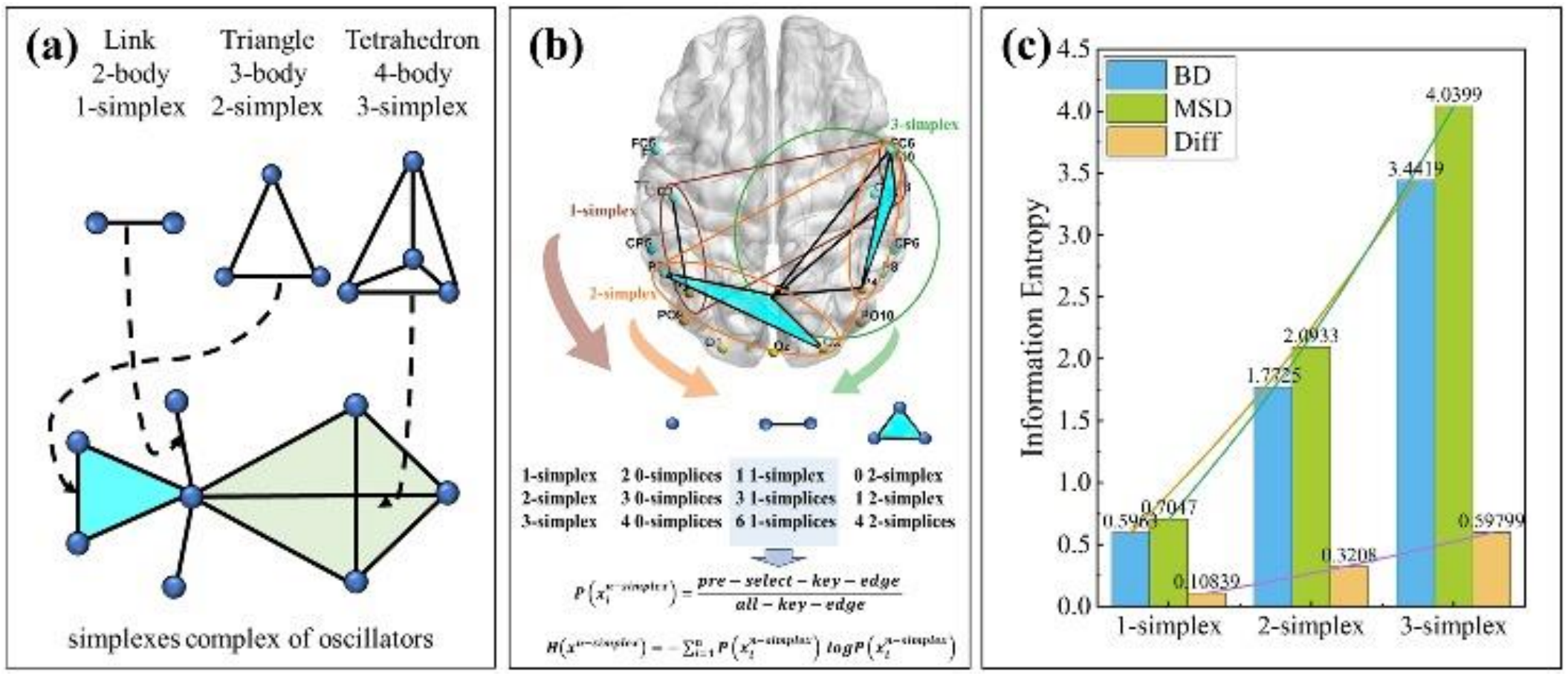

**Fig. 5.** High-order network analysis of brain states during BD and MD stages. (a) Illustration of high-order network structures, including binary interactions (links), ternary interactions (triangles), and quaternary interactions (tetrahedrons) within a simplicial complex. (b) High-order network constructed from ROI-based brain regions. (c) Comparative analysis of information entropy across different orders in high-order networks between BD and MD stages and their difference matrix *diff* statistics.

### *C. Interaction Entropy-based Higher-Order Brain Network Analysis*

Higher-order networks represent a mathematical framework for modeling interdependencies among multiple elements in complex systems, capturing structural and functional patterns that conventional network models cannot encapsulate. These networks are particularly significant in brain network analysis, as they provide a more nuanced characterization of neural and functional connections, offering deeper insights into the organizational principles of complex systems. To systematically analyze the evolution of brain networks, quantify the connectivity diversity across networks of varying orders, and evaluate information integration during cognitive tasks, this study employs interaction entropy and integrated information capacity, denoted by $\Phi$. Both measures are directly linked to the brain's information processing and integration capabilities. Interaction entropy quantifies the global irregularity of neural activity patterns and the brain's information processing capacity, emphasizing the quantity and diversity of information that can be processed. In contrast, $\Phi$ specifically measures the degree of information integration within the brain system, focusing on the extent of synergy and information exchange among its constituent parts.

As illustrated in Fig. 5(a), binary networks, termed first-order networks or 1-simplices, represent the most fundamental network form, consisting of an edge connecting two brain nodes and indicating functional associations between brain regions. These networks provide foundational insights into basic connection patterns in the brain. Ternary networks, or second-order networks (2-simplices), extend this concept by considering three-node relationships, often forming triangles that reflect broader patterns of synergy and enhance the efficiency and accuracy of information transmission. Quaternary networks, or third-order networks (3-simplices), involve interactions among four nodes, typically represented by tetrahedrons. By utilizing simplices to model node interactions in higher-order networks and constructing simplex networks based on key edges and their associated nodes, this approach enables a more comprehensive understanding of the brain's information integration mechanisms during complex cognitive tasks.

In this study, based on 34 key edges within a ROI comprising 20 nodes, we examined the third-order network structure. The 3-simplicial complex contains 4 0-simplices, 6 1-simplices, and 4 2-simplices. As illustrated in Fig. 5(b), we calculated the joint distribution of 1-simplices across different network orders to determine the interaction entropy. Fig. 5(c) demonstrates that interaction entropy increases with network order across two stages, primarily because higher network orders reflect the joint distribution properties of multi-regional node combinations, necessitating more complex coordination mechanisms due to the exponential growth in the number of node groups requiring coordination.

During the MSD stage, phase locking between auditory and motor rhythms generates novel ternary combination patterns, enabling dynamic coordination between AN and SMN and establishing auditory-motor resonance. Concurrently, visual navigation information integrated with SMN motion planning via DMN scene memory forms a four-node coordination. Music-induced motor imagery and real-time visual navigation create predictive coupling, enhancing visual-motor conversion efficiency and promoting 'VN-DMN-SMN-AN' cross-modal integration. Furthermore, DMN engages in self-referential motion strategy selection and establishes bidirectional regulation with SMN.

Notably, under musical stimulation, dynamic coordination, cross-modal integration, and indirect compensation between functional regions are significantly enhanced, resulting in higher interaction entropy during the MSD stage compared to the BD stage under the same network order. Importantly, the largest difference in interaction entropy between MSD and BD occurs at the third-order network, reflecting music-enhanced coordination among quaternary components. Task complexity necessitates network reconfiguration, positioning the third- order network in the MSD stage as a key regulator of inter-regional coordination, thereby substantially improving global efficiency.

Interaction entropy underpins the computation of Φ values. High entropy levels in the brain indicate rich neural states and advanced information processing capabilities, providing the foundation for information integration and potentially leading to higher Φ values associated with more complex conscious experiences. The existence of Φ values highlights the brain's strategic utilization and organization of entropy during information integration, where scattered information is rendered more meaningful through inter-regional interactions, thereby influencing the distribution and dynamics of brain entropy.

As illustrated in Fig. 6(a), the connectivity patterns of first-order (line), second-order (surface), and third-order (volume) networks within ROI regions during the BD and MSD stages exhibit distinct topological properties. Relative to the BD stage, the MSD stage demonstrates enhanced network characteristics, including increased edge count, progression toward a fully connected topology, reduced shortest path lengths, and accelerated information transmission. These advancements establish a structural foundation for second-order networks while introducing "structural holes" that facilitate localized information processing. The formation of triangles within the network provides three redundant pathways, thereby enhancing local robustness and improving information reuse through nonlinear gain.

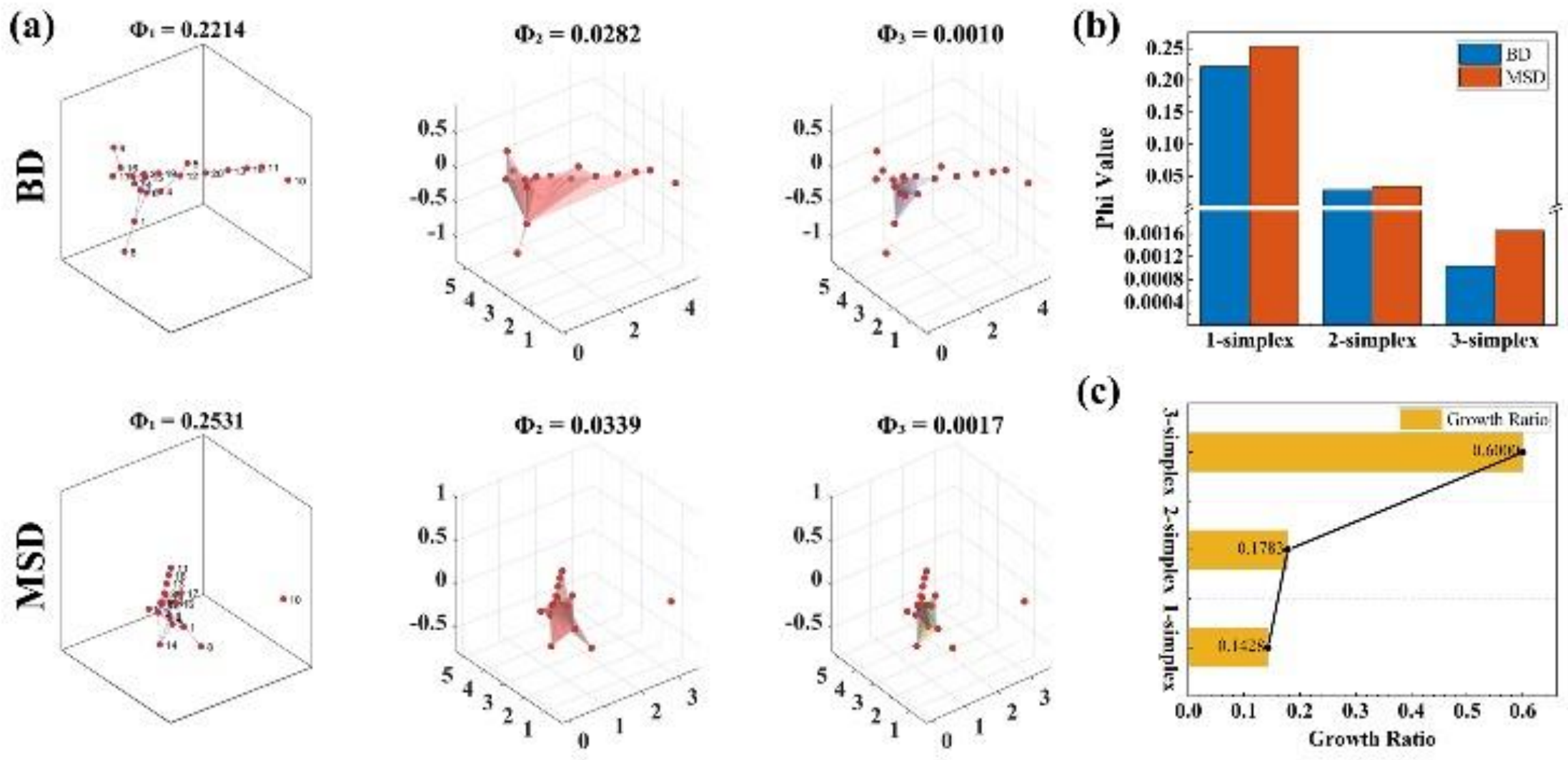

**Fig. 6.** High-level network mutual information analysis. (a) Connectivity matrix of the high-level network constructed from ROI regions during BD and MSD phases. (b) Comparison bar plot of Φ values in the high-level network between BD and MSD phases. (c) Line plot illustrating the percentage change in Φ values within the high-level network from BD to MSD phase.

The elevated density observed during the MSD stage underscores enhanced local integration, supporting multi-dimensional information synergy and promoting the development of a "tetrahedron" structure across multiple regions. Notably, the increased density of body structures reflects improved global integration capabilities. As shown in Fig. 6(b), Φ values for networks across all orders are significantly higher during the MSD stage compared to the BD stage. The densification of critical edges reduces average path length, thereby enhancing basic network performance. Additionally, music strengthens the association between visual scenes and self-referential memory, improves the efficiency of converting environmental visual information into motor control, and balances driving automation with consciousness monitoring. This facilitates the construction of the "AN-VN-SMN-DMN" tetrahedron, establishing an optimal loop of "auditory information-visual scene mapping-motor execution-self-monitoring" and increasing inter-regional synergy efficiency, as evidenced by elevated $\Phi$ values.

Importantly, as shown in Fig. 6(c), the growth rate of Φ values in the third-order network during the MSD stage surpasses that of first- and second-order networks. This indicates that, under musical influence, the third-order network—critical for addressing complex tasks—enhances global integration and strengthens predictive control, thereby optimizing structural integration and resource allocation in the brain.

### *D. Higher-Order Network Metrics as Predictors of Motor-Control States*

To investigate the impact of environmental inputs on behavioral outputs, Fig. 7 illustrates a closed-loop information processing framework. Environmental changes are processed through a perception module, which captures visual, auditory, and sensory cortical information. These inputs are subsequently analyzed by the cortical analysis module, where high-order brain networks (HOBN) generate node combinations and extract relevant coefficients. The system then constructs multi-dimensional samples and applies multi-dimensional binning to compute the joint probability distribution. Using this distribution, the joint information index is calculated and processed through machine learning methods, ultimately influencing the behavioral output HDA. This establishes a comprehensive, closed-loop workflow linking environmental inputs to behavioral metric outputs through integrated information processing and decision-making mechanisms.

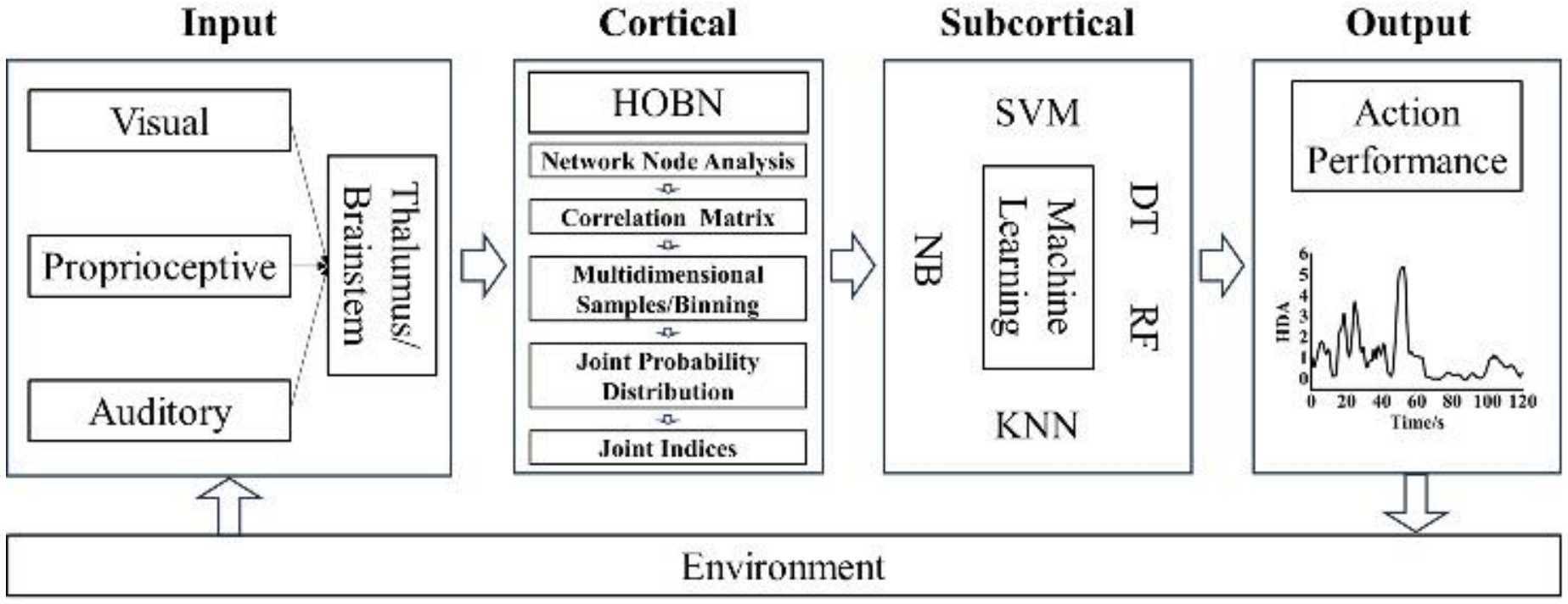

**Fig. 7.** Architecture diagram of brain-environment interaction for motor-control mode prediction during simulated driving. This diagram illustrates the interaction mechanisms between the brain and a simulated driving environment. Sensory information (visual/auditory and somatosensory) is received by the thalamus/brainstem and transmitted to the cortex, where it is primarily encoded as high-order network information. This decoded information is subsequently relayed to the subcortex, which generates motor commands for muscle neurons to execute motor-controlled movements. The subcortical function in classifying and transforming electrical EEG signals into muscle-motor-control mechanical signaling is represented by the block-box MLs. The performance of these movements is quantitatively assessed through the observed behavioral output of HDA levels.

To determine whether higher-order brain network metrics can predict brain states, we employed supervised machine learning models, as illustrated in Fig. 8(a). The models included SVM, KNN, DT, RF, and NB. For predictor design, we extracted node data from DMN, SMN, AN, and VN regions. Phase synchronization matrices between these regions during BD and MSD stages were computed. A threshold $\alpha$ was applied to identify 34 edges with significant changes. We then calculated higher-order information entropy measures, including 1-simplex, 2-simplex, and 3-simplex entropies, for these key edges. These entropy values were used as input features for model training, with $n = 15$ samples, and predictive accuracy was compared across different orders.

As shown in Fig. 8(b), the classification accuracy of all five models achieved optimal performance with third-order features, with SVM, DT, and RF models reaching an accuracy of 0.8889. A notable improvement in accuracy was observed for KNN, DT, and RF models when transitioning from second-order to third-order features. Fig. 8(c-g) illustrates that the ROC curves of the five models demonstrated AUC values around 0.87 for first-order features, increasing to 0.91 for second-order features, and reaching approximately 0.96 for third-order features. These findings suggest that binary connections alone are insufficient for reliable state differentiation due to limited perturbation in the brain's foundational connectivity. The isolated changes in "lines" lack specificity, perceived as weak features by the models.

Music-induced reconfigurations were observed in local functional circuits within the visual-motor-frontal attention cortex and auditory-edge system-motor cortex, reflected by increased second-order entropy. Model classification effectiveness was achieved through identifying the "birth and death" of these closed-loop circuits. At the third-order level, music dynamically transformed the brain's global resource allocation hubs. The original visual-frontal-motor-parietal tetrad was dismantled, and a "music-driven tetrad" (auditory-edge system-motor-default network) was constructed. Models achieved the strongest discriminative power by detecting the topological features of these tetrad hubs. Music does not randomly disturb "lines" but hijacks the cooperative high-dimensional structure to dominate brain network reorganization.

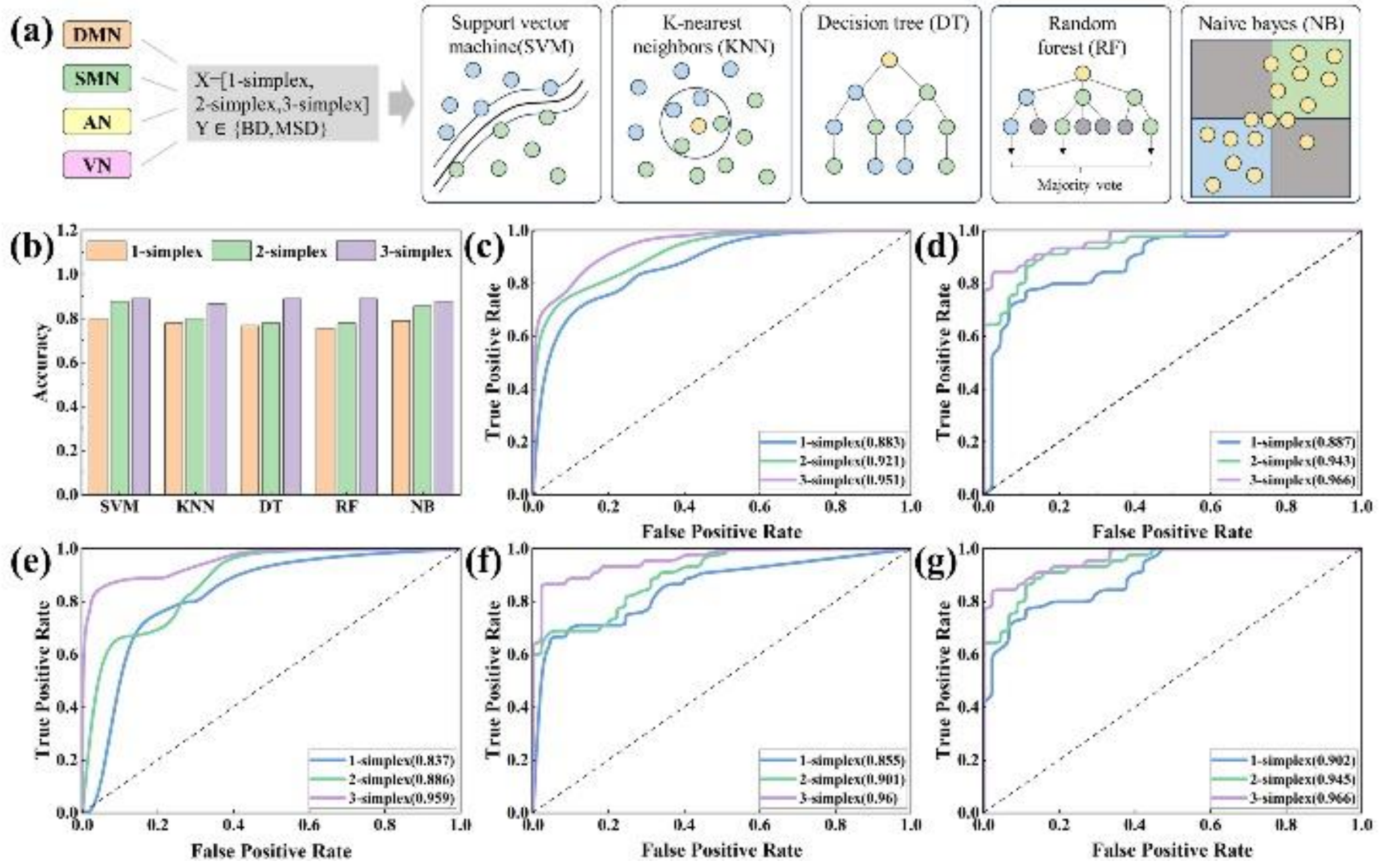

**Fig. 8.** Assessment of high-order metrics' predictive capability through machine learning approaches. (a) Overview of employed machine learning models, including Support Vector Machine (SVM), K-Nearest Neighbor Algorithm (KNN), Decision Tree (DT), Random Forest (RF), and Naive Bayes (NB). (b) Comparative analysis of model accuracy across different order values using four distinct machine learning algorithms. (c-g) Comparative ROC curves illustrating the impact of varying order values on classifier discrimination performance for each model (SVM, KNN, DT, RF, NB, respectively).

This study introduces a novel paradigm for fatigue driving risk prediction. Fatigue driving disrupts high-order network coordination, exhibiting spatial topological isomorphism with music stimulation, primarily through overactivation of the "default network tetrad." The detection of "fatigue topological features" could trigger customized interventions, utilizing rhythmic music stimulation to reconstruct the "auditory-motor" triad and enhance alertness. Simultaneously, this approach activates antagonistic high-order modules while suppressing the overactive "default network tetrad," providing mechanistic intervention targets for fatigue driving and driving active safety systems. This shifts the focus from "behavioral responses" to "neural topological reconstruction," offering a promising direction for advancing driving safety technologies.

## V. Discussion

This study explores stage-specific enhancements in phase synchronization across brain regions. During baseline driving (BD), strengthened phase synchronization between the frontal lobe and central region facilitated motor control, while limbic system activation supported primary information integration. In music-intervened driving (MSD), the frontal lobe retained cognitive control and enhanced functional coupling with the temporal lobe [see Fig. 3(a)], supporting rhythm-based motor coordination [27], [28], [46] and revealing parallel mechanisms for dual-task processing. Additionally, increased synchronization between the parietal lobe and the central region's limbic belt formed a critical circuit for spatial navigation and sensorimotor integration, aligning with the theta oscillation model of cross-modal integration [26], [33]. Brain network phase topological analysis [see Fig. 3(b)] further showed that heightened internal connectivity density in the frontal lobe strengthened modular processing of the cognitive control network, while enhanced frontal-temporal coupling established a direct channel for auditory cognition.

Our findings highlight music stimulation's impact on brain network integration, with a key innovation: identification of a novel higher-order network mechanism beyond conventional binary connections. Three-way hyperedge analysis [see Fig. 4(a)] confirmed strengthened core connections and identified three distinct enhanced three-way patterns: music-binding (AN-SMN), emotion-driven (DMN-SMN), and multimodal integration (VN-DMN/SMN). These address limitations of traditional phase synchronization analysis by revealing a nonlinear information gain mechanism—characterized by increased three-way $\Phi$ values [see Fig. 6(c)]—that transcends linear constraints of conventional second-order connections [33], [34], [47], [48]. Notably, the three-way network's interaction entropy difference peaks during MSD [see Fig. 5(c)], linking to neural coding mechanisms where gamma-band power modulation influences subjective pleasantness [49] and offering new insights into how emotional valuation shapes higher-order integration.

This study integrates dynamic phase synchronization theory with higher-order network analysis to propose a hierarchical model with three stages: (1) establishing multimodal channels via visual-motor network interactions (VN-SMN); (2) forming a visual-motor triangular structure to boost local efficiency; and (3) constructing a tetrahedral AN-VN-SMN-DMN collaborative structure for global resource allocation. This framework explains how music optimizes driving decisions by enhancing higher-order coordination among visual, auditory, and sensorimotor networks. Synthesizing prior findings [26], [28], [33], [49], it provides a unified understanding of brain network reorganization under complex tasks. Validation via supervised machine learning—showing classification accuracy [see Fig. 8(b)] and ROC-AUC values [see Fig. 8(c-g)] correlate with structural order—highlights higher-order spatial coordination as critical for decoding music's influence. Furthermore, while fatigued driving impairs higher-order coordination, music intervenes via topological reconstruction, offering novel pathways for fatigue detection and mitigation. This work aims to lay theoretical groundwork for "brain network spatial architecture-based" driving warning systems.

Limitations remain. First, dynamic implications of reduced phase synchrony—potentially reflecting excitatory-inhibitory balance—are unexplored. Though prior work links alpha oscillation phase resetting to local inhibitory activity [50], synchrony decay in specific connections here may represent active inhibition of task-irrelevant circuits, critical for understanding resource allocation. Second, limited sample size constrains higher-order network analysis power; as Bassett et al. [42] note, such analysis requires robust statistics, with individual variability potentially obscuring subtle higher-order topological effects. Third, cross-frequency coupling's interaction with higher-order topology is unexamined: theta-nested gamma oscillations may modulate integration via modular structure, and beta-band power-phase coupling may drive edge remodeling—mechanisms uncaptured in the current single-frequency framework. Future work should develop multi-frequency super-network models, integrate time-frequency and topological analyses, investigate dynamic inhibitory regulation, and expand samples to enhance detectability. Dynamic causal modeling (DCM) could further quantify cross-frequency coupling's impact on information flow, fully elucidating how musical stimulation dynamically reorganizes cognitive networks across scales.

## VI. Conclusion

Music is known to resonate with brain networks, enhancing cognitive processing. However, the neural mechanisms underlying music-induced improvements in cognitive efficiency during human-machine interaction remain unclear. This study investigates the dynamic neural mechanisms and high-order network properties induced by music stimulation during driving tasks, from high-order brain network perspectives. It examines the predictive capacity of high-order network metrics for brain working modes. Results reveal that music stimulation enhances multimodal visual-motor -auditory information processing efficiency by reconfiguring inter-regional functional connectivity patterns. This improvement is primarily mediated through enhanced cross-level phase synchronization and optimized high-order network coordination. Notably, the machine learning algorithms -based topological features of these high-order coordination structures effectively distinguish brain working modes (cognitive capacity state involving visual-motor function during music-driven driving tasks), offering valuable insights for the development of adaptive human-machine systems.

## Acknowledgement

This work was supported by the National Natural Science Foundation of China (Grant Nos. 81870863,12002251), the Postdoctoral Scientific Research Foundation, General Hospital of Central Theater Command (Program No. 20220224KY29).

## Declaration of Interest Statement

The authors declare that they have no conflict of interest.